\documentclass[a4paper]{article}

\usepackage[margin=1in]{geometry} 

\usepackage{amsmath}
\usepackage{amsthm}
\usepackage{amssymb}
\usepackage{bm}
\usepackage[utf8]{inputenc}
\usepackage{hyperref}

\usepackage[sort&compress,numbers,square]{natbib}
\usepackage{graphicx, color}

\usepackage{algorithm, algpseudocode} 
\usepackage{mathrsfs} 

\usepackage[aboveskip=1pt,labelfont=bf,labelsep=period,justification=raggedright,singlelinecheck=off]{caption}

\usepackage{nameref,hyperref}
\usepackage{array}
\usepackage{float}
\usepackage[caption=false,labelformat=simple]{subfig}

\title{Spatio-temporal spread of COVID-19 and its associations with socioeconomic, demographic and environmental factors in England}
\author{Xueqing Yin$^1$\thanks{Corresponding author: de22002@bristol.ac.uk} \and John M. Aiken$^{2,3}$ \and Richard Harris$^1$ \and Jonathan L. Bamber$^{1,4}$}

\date{
	$^1$School of Geographical Sciences, University of Bristol, Bristol, UK \\
	$^2$   Expert Analytics, Oslo, Norway \\ 
 	$^3$  Njord Centre, Departments of Physics and Geosciences, University of Oslo, Oslo, Norway \\
        $^4$ Department of Aerospace and Geodesy, Technical University of Munich, Munich, Germany
}

\begin{document}
	\maketitle
	
	\begin{abstract}
		Exploring the spatio-temporal variations of COVID-19 transmission and its potential determinants could provide a deeper understanding of the dynamics of disease spread. This study aims to investigate the spatio-temporal spread of COVID-19 infection rate in England, and examine its associations with socioeconomic, demographic and environmental risk factors. Using weekly reported COVID-19 cases from 7 March 2020 to 26 March 2022 at Middle Layer Super Output Area (MSOA) level in mainland England, we developed a Bayesian hierarchical spatio-temporal model to predict the COVID-19 infection rates and investigate the influencing factors. The analysis showed that our model outperformed the ordinary least squares (OLS) and geographically weighted regression (GWR) models in terms of prediction accuracy. The results showed that the spread of COVID-19 infection rates over space and time was heterogeneous. Hotspots of infection rate exhibited inconsistent clustered patterns over time. Among the selected risk factors, the annual household income, unemployment rate, population density, percentage of Caribbean population, percentage of adults aged 45-64 years old, and particulate matter concentrations were found to be positively associated with the COVID-19 infection rate. The findings assist policymakers in developing tailored public health interventions for COVID-19 prevention and control.
		
	\end{abstract}

	
	\section{Introduction}
	\label{sec:intro}

 Coronavirus Disease 2019 (COVID-19) has caused an enormous challenge to global public health, social security and the economy. The disease originated in Wuhan in China in December 2019, and was declared a pandemic by the World Health Organisation (WHO) in March 2020 \cite{who2020world}. As of February 2023, over 650 million people have been infected by the virus across the world, and the global death toll linked to COVID-19 has been recorded at over 6.7 million (\url{https://covid19.who.int/}). Although vaccination and public health measures have been successful in reducing the spread of COVID-19 in many areas, gaining a comprehensive understanding of the spatio-temporal transmission patterns and key determinants of the virus remains crucial in the ongoing public health fight against this global health threat. Recent research studies have proposed that COVID-19 transmission and severity may be influenced by various risk factors. Some researchers examined the association of COVID-19 mortality and infections data with socioeconomic status (e.g., deprivation index, income and unemployment), and found that the virus hit harder in socioeconomically disadvantaged communities \cite{sun2021spatial, choi2021studying, akinwumiju2022geospatial, feng2022spatial}. Some scholars indicated that environmental conditions such as air pollution, humidity and temperature could influence the virus transmission \cite{wu2020air, mecenas2020effects, diaz2020association,wang2020high, berg2021long}. In addition, there have been papers on investigating the demographic influencers (e.g., race, gender, age distribution) of COVID-19 transmission \cite{tamrakar2021district, wong2020spreading, gebhard2020impact,al2021demographic,harris2021measuring,mansour2021sociodemographic,kim2021covid,richardson2022association, green2023occupational}. However the available evidence on these determinants to date presents a mixed and inconsistent picture. For example, Castr et al.\cite{castro2021spatial} found a strong positive relationship of COVID-19 incidence and mortality with income, whereas Liu et al.\cite{liu2021impacts} showed that higher income groups demonstrated lower incidence rates. A multi-country study carried out by Huang et al. \cite{huang2021long} found significant effects of $\text{PM}_{2.5}$ on COVID-19 incidence in the USA but not in Italy or Canada. Furthermore, few studies have accounted for socioeconomic, demographic and environmental factors simultaneously to explain the spatio-temporal variations of COVID-19 infection. Therefore, this paper adds to the global evidence base on the topic of identifying influencing factors of COVID-19 spread, by presenting a new study that investigates the spatio-temporal spread of COVID-19 infection rates and its associations with socioeconomic, demographic and environmental factors in England. Our study is based on aggregated count data summarising COVID-19 infected cases between 7 March 2020 and 26 March 2022 at Middle Layer Super Output Area (MSOA) scale in England and by week. The data are modelled using a Bayesian hierarchical model, where the spatio-temporal variation in COVID-19 infection rate is explained by a set of risk factors of interest, important confounding factors, and random effects that account for the spatio-temporal autocorrelation in the data.

To the best of our knowledge, this study represents the first attempt to incorporate socioeconomic, demographic and environmental factors simultaneously to analyse the spread of COVID-19 infection rates at MSOA level across England. Furthermore, it is one of the most up-to-date and comprehensive studies in terms of its temporal duration. For example, Sun et al. \cite{sun2021spatial} used data up to May 2020, Sartorius et al. \cite{sartorius2021modelling} used data up to 23 August 2020, while the data used by Harris \& Brunsdon \cite{harris2021measuring} goes up to 21 May 2021. In addition, compared to previous studies that mostly used frequentist approaches, such as the geographically weighted regression \cite{castro2021spatial,wu2021exploration, zhang2021space}, spatial error model \cite{kim2021covid, sarkar2021spatial, martines2021detecting} and spatial lag model \cite{kulu2021infection,jackson2021spatial, liu2021spatiotemporal}, to identify the drivers of the spread of COVID-19, the Bayesian hierarchical model developed here has several advantages. Firstly, it allows for the incorporation of prior knowledge and a more robust assessment of the uncertainties in estimates by specifying prior distributions through a hierarchical modelling scheme \cite{gelman1995bayesian, wintle2003use,cressie2009accounting}. Secondly, the model accounts for spatio-temporal autocorrelation in the data by modelling the random effects via a continuous Gaussian random field process. Thus it is able to account for unreported or missing case data by borrowing information from nearby locations in space and/or time based on the modelled spatio-temporal dependence structure. This enables predictions of COVID-19 infection rates in any geographical location and at any time point, even if the infection cases have not been recorded. This is important from a public health perspective, because COVID-19 is often asymptomatic and under-reported globally, especially during the early phase of the pandemic when there was a lack of resources to facilitate large-scale testing \cite{lee2021quantifying,harris2021measuring}. Therefore, the ability of the model to predict the infection rate even in areas where data are incomplete could provide decision-makers with more comprehensive and accurate information to guide interventions and resource allocation in the fight against the pandemic. This study quantifies the spatio-temporal patterns and disparities in COVID-19 infection rates across England, while also identifying hotspot areas and providing additional evidence on the influencing factors of COVID-19 spread. The Bayesian model utilised in this analysis has demonstrated its effectiveness in accurately predicting COVID-19 infection rates, by exhibiting better modelling accuracy than the ordinary least squares (OLS) and geographically weighted regression (GWR) models. The obtained findings provide valuable insights for policymakers to optimise healthcare resources, establish targeted public health interventions, and improve epidemic prevention and control systems.

\section{Materials and methods}

\subsection{Ethics statement}
No specific permissions were required for the described study, because we did not carry out any experiments. All the work in this study was developed using published data from the official UK Government COVID-19 dashboard, Office for National Statistics and Department for the Environment, Food and Rural Affairs.  

\subsection{Study region and data} \label{sec:outcomedata} 
\begin{figure}[H] 
\centering
\captionsetup[subfigure]{justification=centering}
\subfloat[\label{fig:region}]{{\includegraphics[scale=0.45]{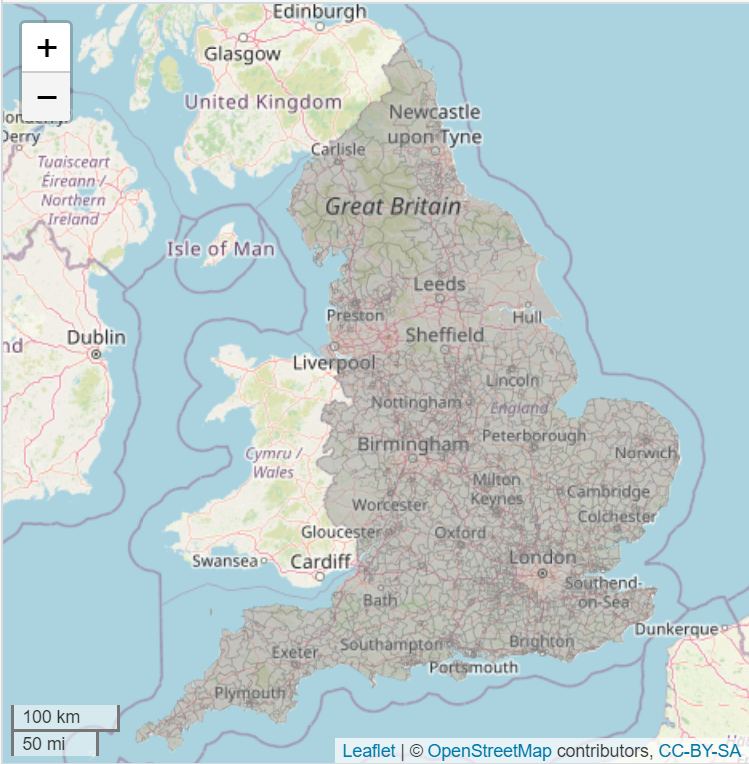}}} \hskip 0.6ex
\subfloat[\label{fig:cases}]{{\includegraphics[scale=0.45]{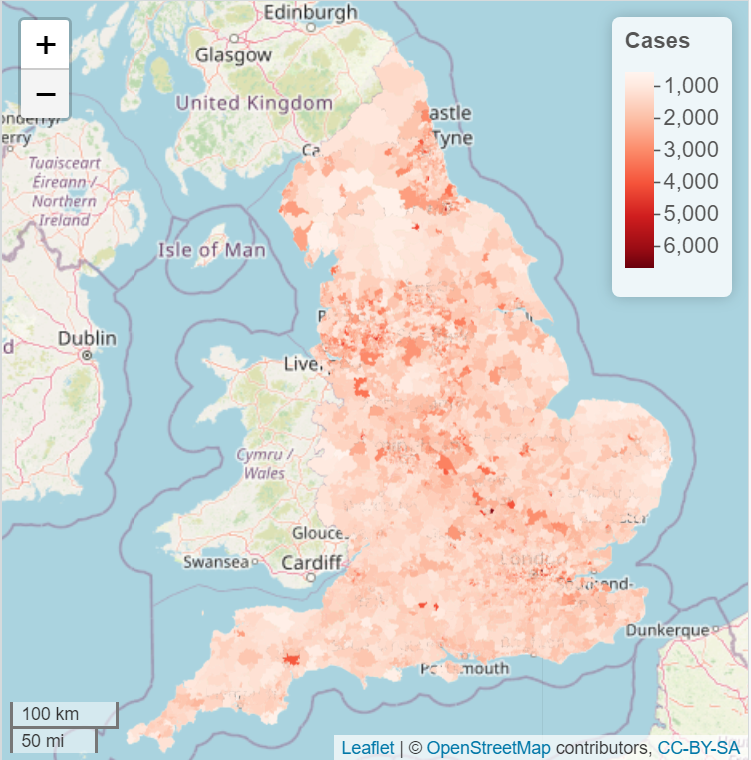}}} 
\vskip 0.4ex
\subfloat[\label{fig:timecases}]{{\includegraphics[scale=0.60]{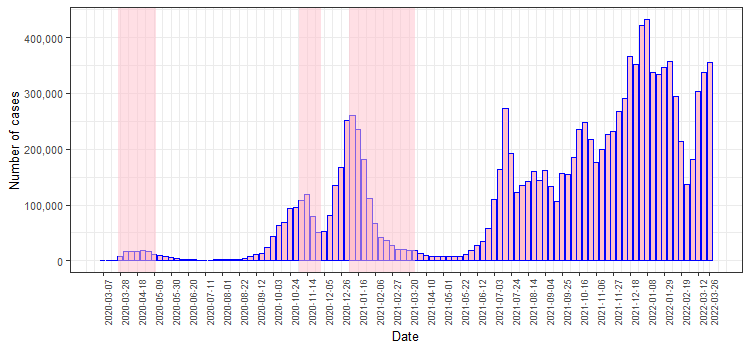}}}
\caption{\textbf{Maps of the study region (a) and cumulative number of COVID-19 infection cases (b) at English MSOA level during the study period. (c): Plot of the reported number of COVID-19 cases by week for all MSOAs in mainland England between 7 March, 2020 and 26 March, 2022. Weeks of lockdown are highlighted in pink.}}
\label{fig:fig1}
\end{figure}

The study region, as shown in Fig~\ref{fig:fig1}(a), is mainland England, which is partitioned into $n= 6789$ small neighbourhood units called Middle Layer Super Output Areas (MSOAs). MSOAs are a statistical geography which are designed by English Government for reporting small-area statistics. They represent a formal, administrative specification of neighbourhoods and are developed with specific criteria, including broadly equal population size, socio-economic similarity based on accommodation type and tenure, and spatial compactness of the zones \cite{cockings2011maintaining}. The shapefiles for MSOAs were obtained from the open geoportal platform (\url{https://geoportal.statistics.gov.uk/}) provided by the UK government. The median geographical size of an MSOA in England is 3.04 $\text{km}^2$, and as of the end of June 2020, the population size in a single MSOA ranges between 4843 and 27911 persons, with a median population size of 8123.

The COVID-19 infection data used for this study were obtained from the official UK Government COVID-19 dashboard (\url{https://coronavirus.data.gov.uk/details/download}) on the weekly number of reported COVID-19 cases for each MSOA. The infections data tally, by date reported, the number of individuals who had at least one positive COVID-19 test result, over that and the preceding six days. The observed time frame of the study spanned $T=108$ weeks from the seven days leading up to 7 March 2020 (week 1) to the seven days to 26 March 2022 (week 108). The data only include pillar 1 cases until 2 July 2020, from when pillar 2 cases are also included. Pillar 1 cases come from the swab testing carried out in Public Health England lab and NHS hospitals for those with a clinical need, and health and care workers, while pillar 2 cases are also swab tests but done for the wider population. These tests are done at local and regional test sites, mobile testing units, satellite test centres and via home tests (\href{https://www.gov.uk/government/publications/nhs-test-and-trace-statistics-england-methodology/nhs-test-and-trace-statistics-england-methodology}{https://www.gov.uk/government/publications/nhs-test-and-trace-statistics-england-methodology/}). A total of 12 097 525 cases (MSOA and week combinations) were reported during the entire study period, whose values ranged from 3 to 978 cases in a single MSOA and week with a median of 28 cases. Note that the positive cases per MSOA range from 3 upwards is because of censoring of smaller numbers (between 0 and 2 cases) in the reported data. Fig~\ref{fig:fig1}(b) displays the spatial pattern of the cumulative COVID-19 infection cases over all weeks at English MSOA level, with higher numbers of infected cases in and around major metropolitan areas such as Newcastle, Manchester, Liverpool, Birmingham, Leicester, Sheffield, and London. Fig~\ref{fig:fig1}(c) presents the weekly number of reported COVID-19 cases for all MSOAs, with the x-axis labeled “date” indicating the start date of each observation week. During the COVID-19 pandemic, the UK government introduced various measures, such as the tiered system, the localise and nationwide lockdowns to limit the spread of COVID-19. Our observation window encompasses three separate periods of national lockdown in England, which were officially announced and took place from 26 March to 12 May 2020, 5 November to 2 December 2020, and 5 January to 28 March 2021, respectively \cite{muegge2023national}. To aid in visualising these periods, they have been shaded in pink in the figure. The figure shows that the first wave of the COVID-19 outbreak in England started in March 2020 and ended at the end of May 2020, whereas the second wave started at the beginning of September 2020 and then reached an initial peak in mid-November. After that the infection levels declined before drastically rising again in late December of 2020 and peaking in early January 2021. There were signs of decrease in the week starting 9 January 2021 until early June, after which the infections started to rise and then wavered until the end of March 2022, with the highest numbers of confirmed cases ($>$ 400 000 cases) observed in early January 2022 (i.e., the week beginning 8 January 2022). Overall, there was typically a reduction in the number of infection cases after the national lockdowns were in place, particularly during the second and third lockdown periods.

A variety of socioeconomic, demographic and environmental factors were selected as potential explanatory variables for the spread of COVID-19 infections, and all these variables were retrieved or prepared at MSOA level. Based on existing studies and data availability, our socioeconomic variables include annual household income, which was available for 2018 (\url{https://www.ons.gov.uk/employmentandlabourmarket/peopleinwork/earningsandworkinghours}), and unemployment rate, which was measured as the percent of people aged 16 years and over unemployed in the labour force, and was available from the 2021 UK census (\url{https://www.nomisweb.co.uk/census/2021/bulk}) held by the Office for National Statistics (ONS). The demographic factors of population density, ethnic and age distributions by MSOA were also available from the 2021 UK census. Population density was measured as the number of persons per square kilometer, and it was found to be highly skewed to the right. Thus a log-transformation was applied. Environmental variables include the annual mean concentrations of fine particulate matter ($\text{PM}_{2.5}$), coarse particulate matter ($\text{PM}_{10}$) and nitrogen dioxide ($\text{NO}_{2}$) for 2021, all of which were measured in \textmu g $\text{m}^{-3}$. We used concentrations data for 2021 rather than for 2022 or 2020, because the data for 2022 were not available, while the data for 2020 were artificially lower than normal concentration levels due to the implementation of lockdowns and mobility restrictions across England in 2020, which significantly reduced transport usage and hence pollution concentrations of that year. Department for the Environment, Food and Rural Affairs (DEFRA) offers 1 $\times$ 1 km gridded annual mean $\text{PM}_{2.5}$, $\text{PM}_{10}$ and $\text{NO}_{2}$ concentrations (\url{https://uk-air.defra.gov.uk/data/pcm-data}), which is spatially misaligned with the irregularly shaped MSOAs. Thus we adopted the simple averaging method \cite{haining2010inference} to obtain the concentrations by MSOA. Specifically, each 1 $\text{km}^2$ gridded concentration has an associated centroid, and the pollution concentration in an MSOA is computed by averaging the grid square concentrations whose centroids are located within the MSOA. If an MSOA does not contain a grid square centroid, it is assigned the pollution concentration from the nearest grid square. Table~\ref{table:covar} lists the variables together with their source of data. Finally, we also considered two other confounding factors that may impact COVID-19 ill health, including the number of care home beds per adult population (\url{https://www.cqc.org.uk/}), and a binary variable indicating whether there is a hospital with emergency facilities or not in an MSOA (\url{https://www.nhs.uk/}).

\begin{table}[H]	
\small\sf\centering
\captionsetup{justification=justified}
\caption{\textbf{Summary of the socioeconomic, demographic and environmental variables used in this study.}} \label{table:covar}
\begin{tabular}{l l l c}  
\hline
\textbf{Theme}&\textbf{Variable} &\textbf{Year}&\textbf{Source}\\
\hline
Socioeconomic& Annual household income (in thousand, £) &2018 & ONS \\
& Unemployment rate (\%)&2021  &ONS\\
Demographic & Log(Population density) & 2021& ONS\\
& Percent of Chinese (\%)&2021  &ONS\\
&Percent of Indian (\%)&2021  &ONS\\
&Percent of Pakistani (\%) &2021  &ONS\\
&Percent of Bangladeshis (\%) &2021  &ONS\\
&Percent of African (\%)&2021  &ONS\\
&Percent of Caribbean (\%)& 2021 &ONS\\
&Percent of white British (\%) &2021  & ONS\\
&Percent of age 18-29 (\%) & 2021 & ONS\\
&Percent of age 30-44 (\%)&2021  & ONS\\
&Percent of age 45-64 (\%) & 2021 & ONS\\
&Percent of 65 years old and over (\%) &2021 &ONS\\
Environmental & Annual mean $\text{PM}_{2.5}$ (\textmu g $\text{m}^{-3}$) & 2021&DEFRA\\
 & Annual mean $\text{PM}_{10}$ (\textmu g $\text{m}^{-3}$) & 2021&DEFRA\\
&Annual mean $\text{NO}_2$ (\textmu g $\text{m}^{-3}$) &2021 &DEFRA\\
 \hline
\end{tabular}
\end{table}

\subsection{Statistical analysis}
An infection rate is the probability or risk of an infection in a population, which in epidemiology is defined as the proportion of affected people in a population during a specific time period \cite{rothman2008modern}. In this study, it was calculated as the ratio of the number of reported COVID-19 cases to the total population in each MSOA and week. However, naively using such infection rate as a measure of the variations in COVID-19 transmission ignores the spatio-temporal autocorrelation that characterises the disease dynamics. Moreover, it also ignores the potential effects of risk factors on COVID-19 infections. Therefore, it is necessary to develop a model-based approach that can capture the spatio-temporal variations of the disease spread, separate the variations from random noise and account for the spatio-temporal autocorrelation structure in the data.

\subsubsection{Spatial and temporal autocorrelation analysis}
The presence of positive spatial autocorrelation in the COVID-19 infections data was evidenced by performing the Moran's I test \cite{moran1950notes}. Moran's I is widely used to measure the level of spatial autocorrelation between adjacent locations globally. The computed Moran’s I value was 0.242 and the associated \emph{P} value was less than 0.0001, indicating statistically significant spatial correlation in the COVID-19 spread throughout England. To assess the temporal autocorrelation in the data, a Ljung-Box test \cite{ljung1978measure} was conducted. The test showed strong evidence of positive temporal correlation in the COVID-19 infection rates over time, as indicated by a \emph{P} value of less than 0.0001.

\subsubsection{Variable selection}
Before modelling the data, we examined the multicollinearity among the variables described in Table~\ref{table:covar} as well as the confounding risk factors using a Pearson's correlation analysis, because the presence of multicollinearity among independent variables can cause overfitting and less reliable inferences about the associations between the response and predictor variables. The air quality components $\text{PM}_{10}$ and $\text{NO}_{2}$ were found to be strongly correlated with $\text{PM}_{2.5}$, with the Pearson’s correlation coefficients of 0.86 and 0.85, respectively. Given that previous studies have identified $\text{PM}_{2.5}$ as an important risk factor for COVID-19 infections \cite{berg2021long, lee2022quantifying, diaz2020association}, we selected $\text{PM}_{2.5}$ over the other two air pollutants for further analysis. The percent of each MSOA’s population who were white British was highly correlated with the percent of African and Indian population, with the correlation coefficients of -0.7 and -0.6 respectively, and hence was not included in the analysis. The percent of the population between 30 and 44 years old was also not considered in the model due to its high correlations with the percent of age groups 45-64, and 65 years and over, with the correlation coefficients of -0.6 and -0.8, respectively. Simultaneously, the variance inflation factor (VIF) was used to verify multicollinearity. The VIF values for all the filtered variables were less than 5 (between 1.02 and 4.89), indicating no serious multicollinearity exists \cite{menard2002applied}. This means that all the predictor variables in the final model were not highly correlated to each other.

\subsubsection{Bayesian spatio-temporal modelling}\label{sec:modelling}

We employed a Bayesian hierarchical model to predict the COVID-19 infection rates at MSOA level over space and time and identify the associated risk drivers, by utilising a generalised linear mixed model with a combination of the selected risk factors, and random effects which account for any residual spatio-temporal dependence of COVID-19 transmission. Let $Y({\bm {s}}_{i},t)$ and $N({\bm {s}}_{i},t)$ be the number of reported COVID-19 cases and total population in MSOA $i\in (1, \ldots, n_t)$ during week $t \in (1,\ldots,T)$, respectively. Here ${\bm {s}}_{i} \in R^2$ denotes the geographical location for MSOA $i$, and $n_t$ represents the number of MSOAs that have reported COVID-19 cases during week $t$. Let $\theta({\bm s}_{i},t)$ be the logarithm of the observed infection rate in MSOA $i$ during week $t$, then it is defined as

\begin{align}
    \theta({\bm s}_{i},t)&=\log\left(\frac{Y({\bm {s}}_{i},t)}{ N({\bm {s}}_{i},t)}\right).
\end{align}

We chose to model the logarithm of the infection rates ($\theta({\bm s}_{i},t)$), because the raw infection rates ($\frac{Y({\bm {s}}_{i},t)}{ N({\bm {s}}_{i},t)}$) were highly skewed to the right. Moreover, researchers \cite{takagi2021meta,leffler2020association} have demonstrated that the logarithmic transformation is a superior choice over probit or logit transformations for explaining the spread of COVID-19. The first level of the Bayesian hierarchical model is the Gaussian specification given by
\begin{align}
\nonumber  \theta({\bm s}_{i},t) &\sim \text{N}(\mu({\bm s}_{i},t), \sigma_{e}^2),\ \  i=1,\ldots,n_t;\ \  t=1,\ldots,T,\\
 \mu({\bm s}_{i},t)&=\bm x({\bm {s}}_{i})^{\top} \bm\beta+\xi({\bm {s}}_{i},t), \label{proposedmodel}
\end{align}  
where $\mu({\bm s}_{i},t)$ is the COVID-19 infection rate in MSOA $i$ and week $t$ on the log scale, and $\sigma_{e}^2$ is the variance of the measurement error. The log-infection rate is modelled by two components, the first of which is the vector of $p$ known covariates $\bm x({\bm {s}}_{i})=(1, x_1({\bm {s}}_{i}),\ldots, x_p({\bm {s}}_{i}))$ related to location ${\bm {s}}_{i}$, including an intercept term, with regression parameters $\bm\beta=(\beta_0,\beta_1,\ldots,\beta_p)$. As the temporally varying covariate information is not available, we cannot include this in the model and the regression parameters are assumed unchanged over time. The second component accounting for the variations in $\mu({\bm s}_{i},t)$ is the spatio-temporal random effect $\xi({\bm {s}}_{i},t)$, which represents the realization of a spatio-temporal process for the logarithm of COVID-19 infection rates after covariate adjustment. The spatial correlation is modelled by location specific random effects through a Gaussian random field process, which captures the correlation via a covariance matrix expressed as a function of distance between locations. The temporal correlation is modelled by a first-order autoregressive (AR1) process. Specifically, the second level of the model is given by
\begin{align}
    \xi({\bm {s}}_{j},t)&=\alpha \times \xi({\bm {s}}_{j},t-1)+\omega({\bm {s}}_{j},t), \text{ and }
\xi({\bm {s}}_{j},1)\sim \text{N}\left(0,\frac{\sigma_{\omega}^2}{1-\alpha^2}\right),
\end{align}
where $\alpha$ is a temporal dependence parameter such that $\vert \alpha\vert < 1$. $\bm\omega({\bm {s}}_{j},t)=\left(\omega({\bm {s}}_{1},t), \ldots, \omega({\bm {s}}_{n_t},t)\right)^{\top}$ is a spatial random effect that is
assumed to follow a multivariate Gaussian distribution and have $\bm\omega({\bm {s}}_{j},t) \sim \text{N}(\bm 0_{n_t}, \sigma_{\omega}^2\bm\Sigma_{\omega})$, where $\bm 0_{n_t}$ is a $n_t \times 1$ vector of zeros and $\sigma_{\omega}^2$ is the marginal variance of the spatial process. $\bm\Sigma_{\omega}$ is the $n_t \times n_t$ covariance matrix with elements
$(\bm\Sigma_{\omega})_{ij}=C(\vert\vert\bm s_{i}-\bm s_{j}\vert\vert),$
where $\vert\vert\bm s_{i}-\bm s_{j}\vert\vert$ is the distance between locations $(\bm s_i, \bm s_j)$, and $C(\cdot)$ is the Matern function \cite{williams2006gaussian} given by 
\begin{equation}
   C(\vert\vert\bm s_{i}-\bm s_{j}\vert\vert)= \frac{1}{2^{\nu-1}\Gamma(\nu)}(\kappa||\bm s_i-\bm s_j||)^{\nu}K_{\nu}(\kappa||\bm s_i-\bm s_j||),\label{eq:matern}
\end{equation}
where $K_{\nu}(\cdot)$ is the modified Bessel function of second kind, and $\Gamma(\nu)$ is the Gamma function. $\nu$ is the smoothness parameter of the Matern covariance function, which is fixed at 1. $\kappa$ is a scaling parameter controlling the spatial correlation range $\rho$, which is the distance at which the correlation function has fallen to about 0.13 and is given by $\rho=\sqrt{8\nu}/\kappa$. 

The third level of the model specifies the prior specifications for the model parameters. The regression parameters $\bm\beta=(\beta_0, \beta_1, \ldots, \beta_p)$ were assigned independent weakly informative zero-mean Gaussian prior distributions with a large variance, i.e., $\beta_j \sim  \text{N}(0, 1000)$, to ensure their values are mainly informed by data. Penalised complexity priors \cite{fuglstad2019constructing} were specified for the correlation range $\rho$ and the marginal standard deviation parameter $\sigma_{\omega}$, with $p(\rho<1.5)= 0.5$ and $p(\sigma_{\omega}>1)=0.01$, indicating a 0.5 probability of $\rho$ being smaller than 1.5, and a low probability of $\sigma_{\omega}$ being greater than 1, respectively. The temporal autoregressive parameter $\alpha$ was also assigned a penalised complexity prior, with $p(\alpha>0)=0.9$. Finally, a weakly informative log-gamma prior was specified for $\sigma_{e}^2$, i.e., $\ln(\sigma_{e}^2) \sim \text{log-Gamma}(1, 0.00005)$.

All analyses were conducted using the statistical software R version 4.2.1 \cite{team2013r}. Model inference was implemented using the stochastic partial differential equations (SPDE) method and the Integrated Nested Laplace Approximations (INLA) algorithm via the R-inlabru package (2.5.2) \cite{rue2009approximate,lindgren2011explicit,blangiardo2015spatial}. This approach has significant advantages in terms of computational efficiency and accuracy when handling high-resolution spatio-temporal processes and large datasets. The INLA implementation of the SPDE approach approximates a continuous Gaussian random field process with the Matern covariance function by a discretely indexed spatial random process known as a Gaussian Markov random field (GMRF). The GMRF has zero mean and uses a sparse precision matrix, which thus substantially reduces the computational cost in matrix algebra operations compared to using dense covariance matrices \cite{rue2005gaussian}. To represent the Matern field as a GMRF, the SPDE approach discretizes the space by defining a mesh composed of non-intersecting triangles that partition the domain of the study area \cite{lindgren2011explicit}. These triangles allow the spatial autocorrelation between observations to be calculated in the modelling process. Then the INLA algorithm estimates the posterior distribution of the latent Gaussian process and hyperparameters using the Laplace approximation \cite{rue2009approximate}. 

Other R packages including spdep (1.2.5) \cite{spde2022}, ggplot2 (3.3.6) \cite{hadley2016ggplot2}, rgdal (1.5.32) \cite{bivand2019rgdal}, dplyr (1.0.10) \cite{wickham2015dplyr}, RColorBrewer (1.1.3) \cite{neuwirth2014colorbrewer}, leaflet (2.1.1) \cite{cheng2019leaflet}, and stats (4.2.1) have also been used for data analysis and visualisation. Model parameter estimates were summarised in posterior mean, standard deviation (SD) and 95\% credible intervals. A risk factor is considered to have a statistically significant effect if the 95\% credible interval of its estimated regression coefficient does not include zero. The main code to complete this analysis is available from \url{https://github.com/XueqingYin/COVID19.git}.

\section{Results}\label{sec:results}

\subsection{Prediction performance}
The model outlined above was applied to the COVID-19 infections data. For comparison purpose, the OLS and GWR models were also fitted to the same data to observe which one produces the most accurate predictions of COVID-19 infection rate. The OLS model is the most common and widely used global non-spatial regression model, while the GWR model is a spatial model that allows the relationships between the independent and dependent variables to vary by locality. Given the spatial nature of the GWR model, it was applied separately to the COVID-19 infections data for each time period, without considering the temporal autocorrelation in the data. To quantify the prediction accuracy of each model, statistical measures including root mean square error (RMSE) and mean absolute error (MAE) for the infection rate predictions were computed. Their mathematical equations are as follows:

\begin{align}
    \text{RMSE}&=\sqrt{\frac{1}{n_{\text{obs}}}\sum_{t=1}^{T}\sum_{i=1}^{n_t}(\hat{\theta}({\bm s}_{i},t) -\theta({\bm s}_{i},t) )^2},\\
    \text{MAE}&=\frac{1}{n_{\text{obs}}}\sum_{t=1}^{T}\sum_{i=1}^{n_t}\vert\hat{\theta}({\bm s}_{i},t)-\theta({\bm s}_{i},t)\vert,
\end{align}
where $\theta({\bm s}_{i},t)$ and $\hat{\theta}({\bm s}_{i},t)$ represent the observed and predicted values at a give location ${\bm s}_{i}$ and time $t$, and $n_{\text{obs}}$ represents the total number of observations across all time steps.

It can be seen from Table~\ref{table:preacc} that our Bayesian model performed best in terms of prediction accuracy.
The RMSE value for the Bayesian model was 72.5\% lower than that of the OLS model and 15.4\% lower than that of the GWR model. The MAE value for the Bayesian model outperformed the OLS model by 74.1\% and the GWR model by 12.5\%. 

\begin{table}[H]	
\small\sf\centering
\captionsetup{justification=justified}
\caption{\textbf{Prediction accuracy of different models for the COVID-19 infection rates.}} \label{table:preacc}
\begin{tabular}{l| l| l}  
\hline
\textbf{Model}&\textbf{RMSE} &\textbf{MAE}\\
\hline
Bayesian &0.0011 & 0.0007\\
\hline
OLS& 0.0040 & 0.0027\\
\hline
GWR& 0.0013 & 0.0008 \\
 \hline
\end{tabular}
\end{table}

Therefore, the subsequent sections will present the primary results obtained from the proposed Bayesian model, because it is the best fitting model in terms of model accuracy. To explore the transmission of COVID-19 infection rate at MSOA level and understand its relationships with socioeconomic, demographic and environmental risk factors, we will focus on the following research questions of interest.
\begin{enumerate}
    \item Are there health inequalities in COVID-19 infection rates among MSOAs in mainland England, and how do these inequalities evolve over time?
    \item What impacts do the socioeconomic, demographic and environmental factors have on COVID-19 infection rate?    
    \item Where are the hotspots of COVID-19 infection rates?
    \item How does the COVID-19 infection rate change over time in different regions of England? 
\end{enumerate}

 \subsection{Overall health inequalities and temporal evolution of COVID-19 infection rate}
Table~\ref{table:sumhyper} provides a summary of the estimates for the spatial correlation range parameter $\rho$ and temporal dependence parameter $\alpha$ from the model. It indicates that the spread of COVID-19 infections in England had a positive spatial and temporal autocorrelation, with $\rho=0.568$ (SD = 0.004, 95\% CI = 0.559-0.575) and $\alpha=0.908$ (SD = 0.003, 95\% CI = 0.905-0.913). These estimates lend further support to the choice of a spatio-temporal model as an appropriate framework for modelling the spread of COVID-19 infection rates.

\begin{table}[H]	
\small\sf\centering
\captionsetup{justification=justified}
\caption{\textbf{Summary of the spatial correlation range parameter $\rho$ and the temporal dependence parameter $\alpha$.}} \label{table:sumhyper}
\begin{tabular}{l c c c c}  
\hline
\textbf{Parameter}&\textbf{mean} &\textbf{SD}&\textbf{0.025quant}&\textbf{0.975quant}\\
\hline
Spatial range $\rho$ &0.568 &0.004& 0.559  &0.575\\
Temporal dependence $\alpha$ & 0.908 &0.003 &0.905 &0.913 \\
 \hline
\end{tabular}
\end{table}

\begin{figure}[H]  
\centering
\captionsetup[subfigure]{justification=centering}
\includegraphics[scale=0.6]{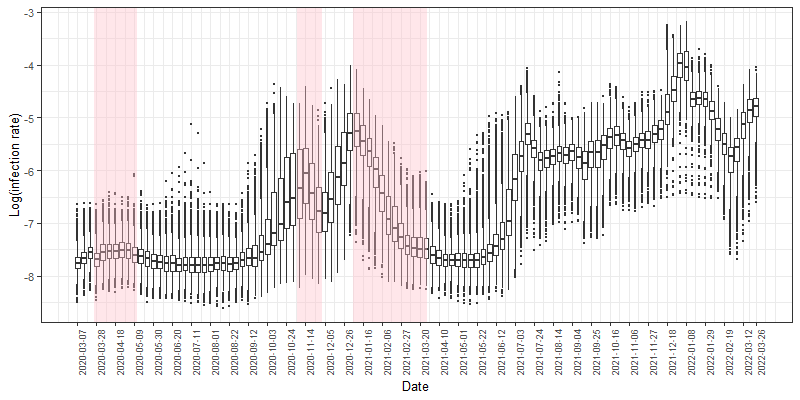}
\includegraphics[scale=0.6]{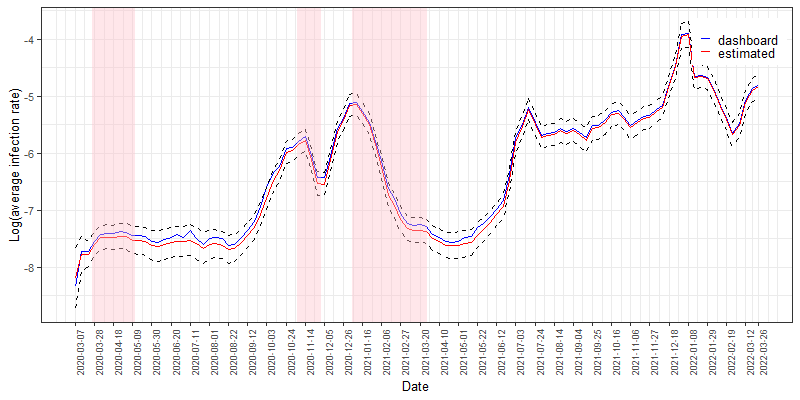}
\caption{\textbf{Summary of the estimated COVID-19 infection rates. The top panel displays boxplots of the logarithm of the estimated infection rates across all MSOAs over the 108 weeks, while the bottom panel compares the model's average estimated infection rate on the log scale across mainland England (with 95\% credible intervals represented by dashed lines) to the log of the average infection rate computed based on UK COVID-19 dashboard reported cases in weeks. The x-axis label "date" indicates the starting date of each observation week. Weeks of lockdown are highlighted in pink.}
}\label{fig:summaryfitted}
\end{figure}

Fig~\ref{fig:summaryfitted} shows two panels illustrating the estimated COVID-19 infection rates for all MSOAs by week. To improve the data visibility, we chose to plot the logarithm of the infection rates. The top panel reveals the presence of health inequalities in COVID-19 infection rates in England. There were substantial variations in the estimated infection rates and the extent of these spatial inequalities increased over time, because the interquartile range of the infection rate, which measures the spread of the distribution, widened from 0.0001 in the first observation week (starting on 7 March 2020) to 0.0028 in the last observation week (starting on 26 March 2022), indicating a potentially growing disparity in COVID-19 infection rates across different neighbourhoods over time.

The bottom panel in Fig~\ref{fig:summaryfitted} shows that the average infection rates estimated by our model were highly consistent with the UK COVID-19 dashboard reported data. Over the study period, the COVID-19 infection rate displayed a series of fluctuations, but there was an overall upward temporal trend. The estimated infection rates ranged from 0.0002 to 0.0417, with a mean of 0.0028 (SD = 0.0037). The average infection rate curve shows an initial peak in mid-November 2020, followed by a decline in infection levels, which then surged in December 2020 due to the Alpha (also known as B.1.1.7) variant of coronavirus. Infections peaked in early January 2021, and then tapered off to almost baseline levels by April of the same year. From the week of 5 June 2021 to mid-July, the infection rate began to rise again, reaching an estimated 0.5\%. The rate then fell and displayed a relatively stable trend with slight oscillations until November 2021. However, the infection rate reached its highest point in the week of 1 January 2022 at 1.9\%, before substantially decreasing until early March 2022. The curve also suggests that the national lockdowns reduced the infection rate, particularly during the second and third lockdown periods, while individuals were at a higher risk of COVID infection during the months of July, December, and January. We also noted that in the first weeks of the study period both the estimated infection rates and their variations across MSOAs were quite low. This is likely because during the first wave of the pandemic between March and July 2020, COVID-19 testing capacity was strictly limited to priority groups due to a lack of infrastructure for large-scale testing, which means that a large number of infected individuals were not formally diagnosed with the virus. As a result, data on confirmed COVID-19 cases were under-reported and incomplete, thus leading to underestimated infection rates and small variation.

\subsection{Risk factor effects}
Table~\ref{table:riskeffects} presents the results of our Bayesian regression analysis, showing the estimated regression coefficients and relative risks (RR) for each of the selected explanatory variables in relation to COVID-19 infection rates. The estimated relative risks and 95\% credible intervals were computed by exponentially transforming the regression coefficients associated with the variables. Note that the relative risks relate to realistic increases in each variable, which are given in brackets in column 1 of the table.

\begin{table}[H]	
\small\sf\centering
\captionsetup{justification=justified}
\caption{\textbf{Estimated regression coefficients, relative infection risks and 95\% credible intervals for the effects of each risk factor on COVID-19 infection rates. The relative risks relate to realistic increases in each covariate, which are given in brackets in column 1 of the table.
}}\label{table:riskeffects}	
\begin{tabular}{l c c}  
\hline
\textbf{Variable} &\textbf{Regression coefficient}&\textbf{Relative risks (95\% CI)}\\
\hline
Annual household income (£1000) &0.0008 &1.0008 (1.0005, 1.0012) \\
Unemployment rate (1\%) &0.0027  &1.0027 (1.0024, 1.0030)\\
Log(population density) & 0.0145 & 1.0146 (1.0129,1.0164)\\
Percent of Chinese (1\%)& -0.0161  & 0.9840 (0.9823, 0.9858)\\
Percent of Indian (1\%)& -0.0005 & 0.9995 (0.9993, 0.9998)\\
Percent of Pakistani (1\%)& -0.0016  & 0.9984 (0.9982, 0.9990)\\
Percent of Bangladeshis (1\%)&-0.0014  & 0.9986 (0.9982, 0.9990)\\
Percent of African (1\%)& -0.0093 & 0.9907 (0.9901, 0.9913)\\
Percent of Caribbean (1\%)& 0.0022 & 1.0022 (1.0009, 1.0036)\\
Percent of age 18-29 (1\%)& -0.0049 & 0.9951 (0.9947, 0.9954) \\
Percent of age 45-64 (1\%)& 0.0031  & 1.0031 (1.0024, 1.0039) \\
Percent of 65 years old and over (1\%)& -0.0067 & 0.9933 (0.9929, 0.9937)\\
Annual mean $\text{PM}_{2.5}$ (1 \textmu g $\text{m}^{-3}$) &0.0125 & 1.0126 (1.0083, 1.0167)\\
Care home beds  (0.01) & 1.3295 &  1.0134 (1.0121, 1.0147)\\
Emergency facilities (TRUE) &0.0007 & 1.0007 (0.9958, 1.0057) \\
 \hline
\end{tabular}
\end{table}

The table clearly demonstrates that the selected variables significantly contribute to the spatio-temporal variations of COVID-19 infection rate, with the exception of the emergency facilities variable because its 95\% credible intervals include the null risk of 1. One of the main drivers of elevated infection rates is socioeconomic factor. We found that both annual household income and unemployment rate were associated with higher infection rates. Specifically, an increase of £1000 in annual household income within an MSOA was associated with a 0.08\% increased infection rate (RR = 1.0008, 95\% CI = 1.0005-1.0012), while an increase of 1\% in unemployment rate was associated with a 0.27\% increased infection rate (RR = 1.0027, 95\% CI = 1.0024-1.0030). Expectedly, the logarithm of the population density was positively correlated with the infection rate. A 1-unit increase in $\log(\text{population density})$ was found to be associated with a 1.5\% rise in COVID-19 infection rate (RR = 1.0146, 95\% CI = 1.0129-1.0164). We also identified interesting patterns in the associations between ethnicity, age groups, and COVID-19 infection rates. Neighbourhoods with a greater percent of Chinese population tend to have lower rate of infections (RR = 0.9840, 95\% CI = 0.9823-0.9858). The Indian, Pakistani and Bangladeshi ethnic groups had a lower level of infections, with rates decreasing by 0.05\%, 0.16\% and 0.14\%, respectively, for every 1\% increase in the percentage of these groups within an MSOA. The African population was also associated with a decreased rate of COVID-19 infection (RR = 0.9907, 95\% CI = 0.9901-0.9913), whereas the Caribbean population had statistically significantly higher COVID-19 infection rates (RR = 1.0022, 95\% CI = 1.0009-1.0036). 

Table~\ref{table:riskeffects} further reveals that infection rates tend to be lower in MSOAs with higher percentages of adults aged 65 years and older (RR = 0.9933, 95\% CI = 0.9929–0.9937), and those with higher percentages of adults aged 18-29 years (RR = 0.9951, 95\% CI = 0.9947–0.9954). Conversely, the population aged between 45 and 64 years old was positively correlated with infection rates, with a 1\% increase being associated with a 0.31\% higher rate. Additionally, there was a positive relationship between $\text{PM}_{2.5}$ concentrations and COVID-19 infections, with a 1 \textmu g $\text{m}^{-3}$ increase in concentrations associated with between a 0.83\% and a 1.67\% increased rate (RR = 1.0126, 95\% CI = 1.0083-1.0167). Finally, increasing the number of care home beds per adult population (RR = 1.0134, 95\% CI = 1.0121-1.0147) was diagnosed positively linked to higher infection rates.

\subsection{Hotspots of COVID-19 infection rates}
\begin{figure}[H]  
\centering
\captionsetup[subfigure]{justification=centering}
\subfloat[Weeks 1-18.]{{\includegraphics[scale=0.5]{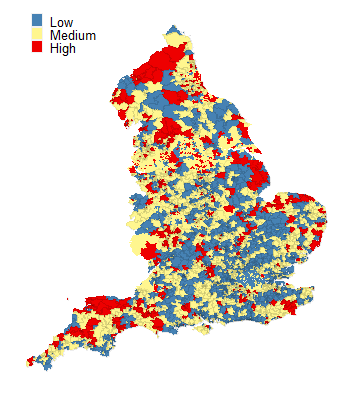}}\label{fig:i1}} \hspace*{-1.1em}
\subfloat[Weeks 19-36.]{{\includegraphics[scale=0.5]{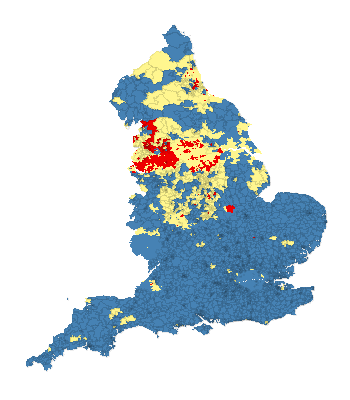}}\label{fig:i2}} \hspace*{-1.1em}
\subfloat[Weeks 37-54.]{{\includegraphics[scale=0.5]{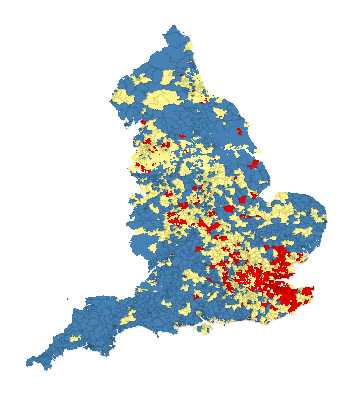}}\label{fig:i3}} 
 \vspace*{-1.3em}
\subfloat[Weeks 55-72.]{{\includegraphics[scale=0.5]{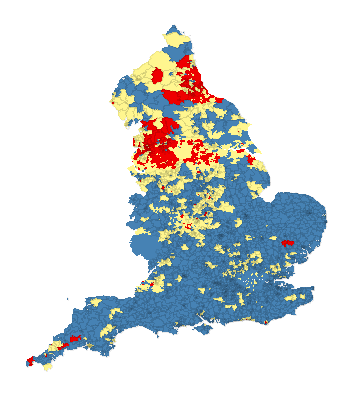}}\label{fig:i4}}\hspace*{-1.1em}
\subfloat[Weeks 73-90.]{{\includegraphics[scale=0.5]{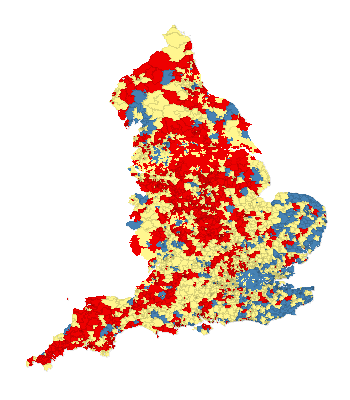}}\label{fig:i5}}\hspace*{-1.1em}
\subfloat[Weeks 91-108.]{{\includegraphics[scale=0.5]{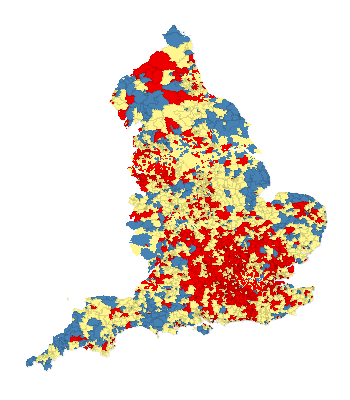}}\label{fig:i6}}
\caption{\textbf{Maps showing the clusters that were formed according to the estimated infection rates at MSOA level during weeks 1 to 18 (7 March 2020-4 July 2020), weeks 19 to 36 (11 July 2020-7 Nov 2020), weeks 37 to 54 (14 Nov 2020-13 March 2021), weeks 55 to 72 (20 March 2021-17 July 2021), weeks 73 to 90 (24 July 2021-20 Nov 2021) and weeks 91 to 108 (27 Nov 2021-26 March 2022), respectively. } 
}\label{fig:hotspots}
\end{figure}

It is of interest to identify which MSOAs showed the high infection rates during the study period, and if there were any areas that consistently exhibited high rates. Clustering methods can be used to identify the groups of MSOAs with high and low rates, and in this study, k-means clustering \cite{hartigan1979algorithm} was utilised. For ease of data visualization, we divided the 108-week time frame into six consecutive and non-overlapping intervals, each spanning 18 weeks. Since k-means clustering uses distance-based measurements to determine the similarity between data observations, we standardised the estimated infections rates during weeks 1 to 18, 19 to 36, 37 to 54, 55 to 72, 73 to 90 and 91 to 108, respectively, and then applied k-means clustering to the standardised rates. The MSOAs were assigned to between one and ten clusters based on their similarities in the estimated rates, and the optimal number of clusters was determined by using the elbow method. Here the optimal number of clusters is 3 clusters for each time interval, and hence the MSOAs were classified into three distinct clusters that have a high, medium, and low level of infection rate, respectively. Fig~\ref{fig:hotspots} displays the spatial patterns of the cluster memberships of MSOAs for each time interval.

Fig~\ref{fig:hotspots} illustrates that the distribution of high infection rates was not consistent across different time periods and exhibited clustering patterns, indicating that certain areas were more significantly impacted by the pandemic than others. During weeks 1 to 18 (7 March 2020-4 July 2020), the highest infection rates were mainly concentrated in northern England, such as districts of Cumbria, Lancashire, Northumberland and Newcastle (County Durham, Gateshead, Sunderland), and in the far southwest, such as Devon. From week 19 to week 36 (11 July 2020-7 Nov 2020), hotspots for COVID-19 infection rates were highly clustered in a contiguous aggregation of MSOAs spanning from Lancashire, Manchester and Liverpool through to Bradford, Leeds, Kirklees and Sheffield. However, this trend moved to the southeast, particularly in London and some surrounding areas such as Kent and Essex in the following 18 weeks. During weeks 55 to 72 (20 March 2021-17 July 2021), the clusters of MSOAs with high infection rates were mainly in and around metropolitan areas for example Manchester, Liverpool, Newcastle, and North Tyneside. Interestingly, weeks 73 to 90 (24 July 2021-20 Nov 2021) showed a more dispersed distribution of hotspots, with hotspots scattered throughout England and no clear clustering patterns observed. However, there was a higher number of hotspots in central England and surrounds (Nottingham, Birmingham, Leicester, Stoke, Coventry, Leeds, Sheffield, Doncaster and Hull), as well as in the southwest (Bath, Bristol, Plymouth and Exeter). Finally, in the weeks 91 to 108 (27 Nov 2021-26 March 2022), high infection rates shifted towards the southeast of England, with the regions of Greater London, South East and East of England most affected. Despite the varying spatial patterns over time, the hotspots tended to cluster in urban and populous areas, particularly in the northwest, central, and southeast. The shifting patterns of infection rates demonstrate the dynamic nature of the pandemic, and the importance of monitoring temporal trends in different regions. Thus in the next section we investigated the temporal trends of COVID-19 infection rates in diverse regions of England.

\subsection{COVID-19 infection rates by region}

\begin{figure}[H]  
\centering
\captionsetup{justification=centering}
\includegraphics[scale=0.55]{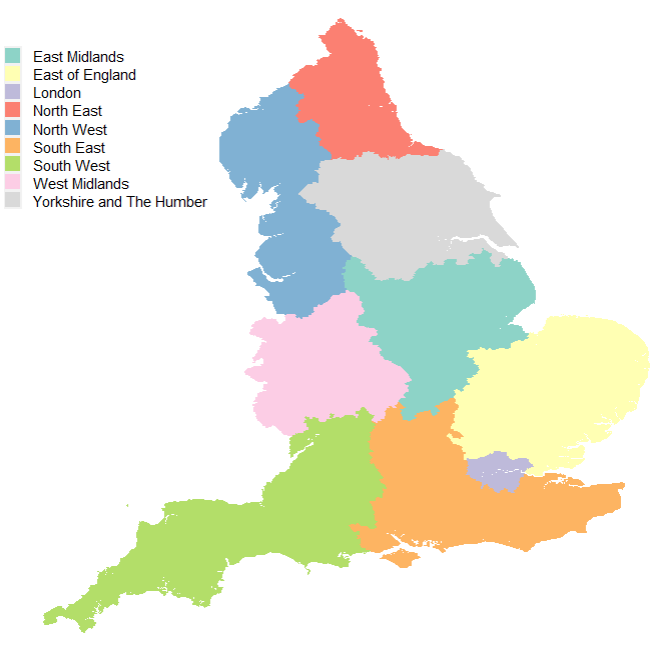}
\caption{\textbf{A map of the nine regions of mainland England.}}\label{fig:rgmap}
\end{figure}

\begin{figure}[H]  
\centering
\captionsetup[subfigure]{justification=centering}
\includegraphics[scale=0.55]{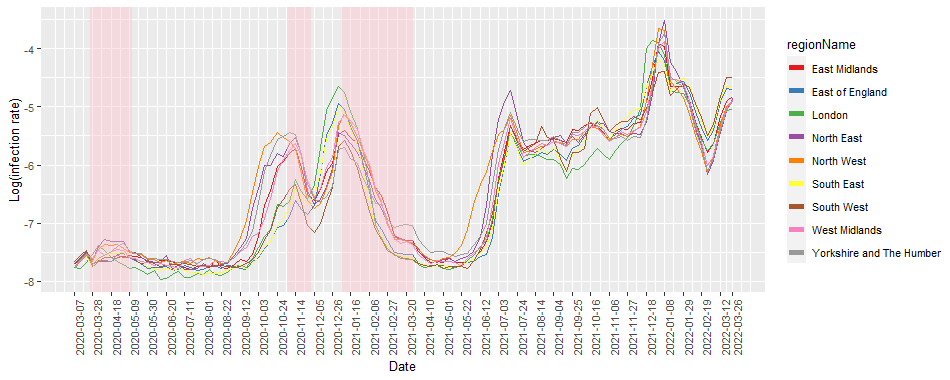}
\caption{\textbf{Population-based weighted averages of COVID-19 infection rates on the log scale by week and region in mainland England. Weeks of lockdown are highlighted in pink.}}\label{fig:rgtrend}
\end{figure}

Differences in COVID-19 infection rate trends can be explored at a larger geographical scale by examining regional data, which provides a more aggregated view of the results. The 6789 MSOAs are nested exactly within nine regions of England, comprising North East, North West, Yorkshire and The Humber, East Midlands, West Midlands, East of England, London, South East and South West. Fig~\ref{fig:rgmap} provides a geographical representation of these nine regions. To characterise the evolution of infection rates over time in each region, we computed the weighted averages of estimated infection rates for region $r$ and week $t$ as $\hat{\phi}_{rt}= \frac{\sum_{i\in r} p_{i}\exp(\hat{\theta}({\bm s}_{i},t))}{p_r}$, where $\hat{\theta}({\bm s}_{i},t)$ denotes the estimated log-infection rate in MSOA $i$ and week $t$, $p_{i}$ and $p_{r}$ indicate the population sizes of MSOA $i$ and region $r$ respectively, and $i \in r$ means that MSOA $i$ is geographically located within region $r$. We used population size as the weighting factor in the computation due to the assumption that neighbourhoods with larger populations are likely to make a greater contribution to the overall infection level of a region compared to those with smaller populations. Fig~\ref{fig:rgtrend} depicts the line plots of the population-based weighted averages of estimated infection rates after logarithm transformation by week and region. Here we chose to plot the log-infection rates instead of the infection rates themselves, as this provides a more suitable scale that allows for better visualisation of the temporal trend curves.

Fig~\ref{fig:rgtrend} suggests that the national lockdown measures have been successful in containing the transmission of the virus, because in all regions the infection rates visibly declined following the initiation of any of the three lockdowns. During the first lockdown period, the Greater London region had the quickest decline in infection rate, while infection rates in other regions, particularly East Midlands, took longer to reduce. After the lockdown was lifted, infection levels remained stable in most regions until August 2020, but increased notably from September 2020 onwards, with peaks observed in mid-November 2020 in all regions except for North West. When the second lockdown was in place, infection rates decreased in all regions, but remained higher in the northern (i.e., North East, North West) and central regions (i.e., Yorkshire and The Humber, East Midlands and West Midlands) of England. After the lockdown was lifted, the rates surged particularly in London, South East, and East of England due to the dominant Alpha variant, which was estimated to be more transmissible than preexisting variants \cite{davies2020estimated}.

During the third lockdown, infection rates across England reduced substantially, with the Greater London region showing the most rapid decline as was the case in the first lockdown. A possible reason for this is that the percentage of key workers in London is relatively lower than that in other regions \cite{batty2021london}. Key workers, such as healthcare workers and public transportation workers, are more likely to contract the virus as they continue to work and interact with the public during the imposition of restrictions. With fewer key workers in London, the lockdown measures may have been more effective and noticeable in reducing virus transmission compared to regions with larger numbers of key workers. There were no lockdown restrictions imposed from April 2021 to the end of the study, during which all regions experienced two infection peaks. One peak occurred in mid-July 2021, with rates ranging between 0.42\% and 0.89\%, while the other was in early Jan 2022, with rates ranging between 1.21\% and 2.61\%. Infection rates remained relatively stable between the peaks, with London having the lowest values. Infection levels sharply reduced from 8 Jan 2022, but rebounded by the end of Feb 2022. South West, South East, and East of England had the highest infection rates throughout the remainder of the study.

\section{Discussion}\label{sec:discussion}

This study aimed to explore the spatio-temporal spread of COVID-19 infection rate in mainland England from 7 March 2020 to 26 March 2022, and its associations with socioeconomic, demographic and environmental factors. It can provide information for public health policies at the neighbourhood level, offering valuable insights for policymakers to optimise healthcare resources, establish targeted public health interventions, and improve epidemic prevention and control systems. Additionally, the study contributes new cases and knowledge to the growing body of research on the space-time transmission of COVID-19.

A Bayesian hierarchical spatio-temporal model was used to predict the COVID-19 infection rates based on weekly reported COVID-19 cases from 7 March 2020 to 26 March 2022 at MSOA level. Note that we utilised the COVID-19 data at MSOA level rather than the other geographical levels such as the Lower Layer Super Output Area (LSOA) level due to data availability. At the time of the analysis, the MSOA level was the smallest geographic unit for which both COVID-19 data and risk factors data were accessible from open sources. We did not downscale the MSOA geographical information to a higher-resolution geographical level, because the downscaling procedure will likely induce bias and inflate the reported credible intervals for the predicted infection rate \cite{konstantinoudis2021long}. Moreover, the MSOA level is commonly used in the literature for studying the small-areas COVID-19 transmissions \cite{sartorius2021modelling}. To evaluate the model's prediction performance, our Bayesian model was compared with the non-spatial regression model (OLS model) and the geographically weighted regression model (GWR model). The RMSE and MAE values of the Bayesian approach were lower than those of OLS and GWR models, indicating that the Bayesian model exhibited better prediction accuracy than the other models in this study.

The model estimation results showed that the level of spatial inequalities of COVID-19 infection rate increased over time, highlighting the need for effective strategies to address the disparities between different neighbourhoods. The COVID-19 infection rates in England exhibited spatio-temporal heterogeneity, with higher rates observed during the months of July, December and January. These findings could be explained by the increased travel, outdoor activities, and social gatherings during the summer months, and holiday celebrations and family gatherings during the winter months. These factors could contribute to higher levels of interpersonal contact within a population, potentially leading to increased transmission risk. Besides, COVID-19 virus may be more transmissible in colder and drier conditions \cite{mecenas2020effects,wang2021impact}, contributing to higher infection rates during the winter months. The infection rates were not evenly distributed across the country, with certain areas more vulnerable to the pandemic. Furthermore, the hotspots of infection rates exhibited clustered patterns that changed over time, with a higher frequency of occurrence in and around urban areas such as Newcastle, Manchester, Birmingham, Liverpool, Nottingham, Sheffield, Leeds, and London. The analysis of regional data indicated that although identical national lockdowns were announced across England, different regions displayed varying impacts in response to these measures. This variation could be influenced by factors such as the proportion of essential workers and the emergence of new variants of the virus. Moreover, it is important to note that different regions in England experienced slight variations in the start and end dates of these national lockdowns due to local epidemiological conditions and the tier system. The tier levels, ranging from Tier 1 (medium alert) to Tier 4 (very high alert), determine the extent of restrictions imposed on social interactions, businesses, and public spaces (\href{https://www.ageuk.org.uk/information-advice/health-wellbeing/conditions-illnesses/coronavirus-guidance/local-lockdown-tiers/}{https://www.ageuk.org.uk/}). Regions with higher infection rates and greater risk were typically assigned to higher tier levels, leading to more stringent measures and restrictions. Thus some regions may have entered lockdown earlier or experienced more prolonged periods of restrictions compared to those in lower tiers. Therefore, the observed differences in response to these measures may also be influenced by the region-specific timing and duration of the lockdown measures.

The analysis of socioeconomic, demographic, and environmental factors in relation to COVID-19 infections further indicated the key factors influencing the COVID-19 landscape in England. We found that MSOAs with higher annual household income had a higher infection rate, which is in agreement with previous investigations conducted in other countries or regions \cite{maiti2021exploring,castro2021spatial}. It is assumed that the influence of annual household income on the infections is related to the presence of a better network of health services, expanding the population's access to carrying out diagnostic tests in the communities with high income, reducing underreporting, and contributing to increased COVID-19 incidence. Another socioeconomic indicator is unemployment rate, which was found to be positively associated with COVID-19 infection rate. Consistent findings have been found to exist in France and the United States \cite{goutte2020role,zhai2021american}. A positive relationship was found between the infection rate and population density. We uncovered that the percent of Asians, including Chinese, Indian, Pakistani, and Bangladeshis, was negatively related to COVID-19 infection rate. This corroborates with the study conducted by Lee et al. \cite{lee2022quantifying}. Conversely, MSOAs with a higher proportion of Caribbean population had elevated infection rates. These relationships may be related to the ethnic differences in COVID-19 vaccine uptake \cite{mathur2021ethnic} and health behaviours (\href{https://www.iser.essex.ac.uk/blog/2021/06/14/are-there-ethnic-differences-in-adherence-to-recommended-health-behaviours-related-to-covid-19}{https://www.iser.essex.ac.uk/blog/}). The age composition of MSOAs was also found to impact COVID-19 infection levels. Higher proportions of adults aged 65 and older were related to lower infection rates. This association could be linked to factors such as higher vaccination rates among older adults, as they were among the first groups prioritised for COVID-19 vaccination in the UK. MSOAs with a higher percentage of young adults aged 18-29 years old showed a lower level of infection, while populations between 45 and 64 years old were positively associated with infection rates. Finally, elevated concentrations of $\text{PM}_{2.5}$ were found to be positively associated with COVID-19 infection rates, which is consistent with existing studies that have reported a positive relationship between increasing $\text{PM}_{2.5}$ concentrations and COVID-19 infections in various regions such as Ohio, Colorado, and Scotland \cite{mendy2021air, berg2021long, lee2022quantifying}. This could be related to the increased susceptibility to respiratory infections such as COVID-19 given exposure to air pollution \cite{fattorini2020role,comunian2020air}. In addition, air pollution has been linked to underlying health conditions such as diabetes, cardiovascular, and respiratory diseases, which are known risk factors for severe COVID-19 \cite{semczuk2021association}. Therefore, it may be necessary to implement more rigorous COVID-19 prevention measures in areas with higher $\text{PM}_{2.5}$.

This study provides evidence that local rates of COVID-19 infections are influenced by patterns of household income, unemployment rate, population density, ethnic composition, age population structure and exposure to air pollution. It provides a scientific basis for accurately predicting COVID-19 response and targeting recovery efforts in England based on community-specific risk factors. It is important to consider these risk factors when developing effective control measures and allocating resources in different communities to control the pandemic spread. For example, public health interventions, such as promoting measures to reduce personal exposure to fine particle pollution could be implemented in specific neighborhoods at higher infection risk, providing better protection for vulnerable populations. Perhaps, most importantly we show that our BHM framework is an effective and powerful tool for modelling and understanding the risk factors that influence the infectious disease dynamics in general. Nevertheless, this study has some limitations. The most significant is its ecological design, where the unit of inference is the group of individuals living in each MSOA rather than having data for each individual. The aggregation of data at the geographic level can cause loss or concealment of certain details about individuals, resulting in ecological fallacy \cite{wakefield2001statistical} in the observed association. Thus, the estimated population-level associations should not be interpreted as cause-and-effect relationships at the individual level, because they may be influenced by factors such as within-area variation in either the exposure or the confounders. We used this ecological study design because the data required for an individual-level study were not available due to confidentiality reasons. We note, however, that this ecological approach has been predominant in the COVID-19 literature that focuses on exploring the distribution pattern of the pandemic and its influencing risk factors \cite{sun2021spatial,nazia2022identifying,mollalo2020gis}. Our findings outlined above should be treated as indicative associations, rather than conclusive evidence of individual-level causation. To fully understand the causal relationships, further research, including tailored experimental designs combined with primary data collection, is necessary to establish causality definitively and investigate the mechanisms and biological pathways behind the observed associations. Secondly, since the annual household income data used were for 2018, the time gap between income data and other predictors and response data may influence the model estimation results. Thirdly, the reported COVID-19 infection cases were linked to the MSOA where the test was conducted, rather than the patient's place of residence. This is a concern for metropolitan areas where a considerable number of patients could have been admitted or transferred to neighboring healthcare facilities before being tested. Finally, the infections data collected in the early phase of the pandemic were often underestimated due to a lack of testing, which could potentially lead to biased estimates and uncertainty in the model predictions. Thus, the limitations of data quality and potential biases should be carefully considered when interpreting the modelling results for the initial pandemic stage.

We noted that in this study the effect sizes of the risk factors on the COVID-19 infection rate were relatively small. The small effect sizes are likely due to the small geographical level (MSOA) and the short time steps (weekly basis) for  reporting the data. MSOAs represent small geographic areas with limited populations, therefore the spatial variations of risk factors and COVID-19 infections between English MSOAs naturally tend to be smaller than those between the larger geographical units such as counties or cities in other studies. Additionally, the use of weekly reporting interval tends to yield smaller infection rate variations compared to longer time steps. However, we believe that even these small effect sizes carry important information and implications for public health strategies. They contribute to our understanding of how specific risk factors influence infection rates at a fine-grained geographical level. By considering these effects, we can gain valuable insights for shaping localised interventions and preventive measures to reduce the spread of COVID-19 at neighbourhood level, and promote community health at this level of granularity. There are several avenues for future work. In this study we assumed fixed effects of risk factors over time, however, we also acknowledged that the associations between these factors and the outcome variable may vary with time, thus there is potential to further enhance the model by incorporating spatially and temporally varying effects of risk factors. In addition, future applications could consider other potentially important predictors for COVID-19 infections such as comorbidities, population mobility and and behavioural factors (e.g., alcohol consumption, smoking). The level of population immunity also plays a crucial role in shaping the disease dynamics and outcomes. To account for this factor in the model, vaccination data could be incorporated as a relevant variable, which is frequently cited as a proxy for population immunity. Lastly, extending the research to other countries and regions could offer comprehensive insights into the global spread of COVID-19 and inform the development of more targeted actions.

	\paragraph{Acknowledgements} 
 The authors thank Dr. Richard Westaway for providing valuable comments on this study.

\bibliography{biblisample}

@misc{who2020world,
  title={World {H}ealth {O}rganization: {C}oronavirus {D}isease 2019 ({COVID}-19) {S}ituation {R}eport},
  author={WHO},
  year={2020},
  publisher={World Health Organization Geneva}
}

@article{lee2022quantifying,
  title={Quantifying the impact of air pollution on {C}ovid-19 hospitalisation and death rates in {S}cotland},
  author={Lee, Duncan and Robertson, Chris and McRae, Carole and Baker, Jessica},
  journal={Spat Spatiotemporal Epidemio},
  volume={42},
  pages={100523},
  year={2022},
  publisher={Elsevier}
}

@article{berg2021long,
  title={Long-term air pollution and other risk factors associated with {COVID-19} at the census tract level in {C}olorado},
  author={Berg, Kevin and Present, Paul Romer and Richardson, Kristy},
  journal={Environ Pollut},
  volume={287},
  pages={117584},
  year={2021},
  publisher={Elsevier}
}

@article{sun2021spatial,
  title={Spatial inequalities of {COVID-19} mortality rate in relation to socioeconomic and environmental factors across {E}ngland},
  author={Sun, Yeran and Hu, Xuke and Xie, Jing},
  journal={Sci Total Environ},
  volume={758},
  pages={143595},
  year={2021},
  publisher={Elsevier}
}

@article{al2021demographic,
  title={Demographic and socioeconomic determinants of {COVID-19} across {O}man-{A} geospatial modelling approach},
  author={Al Kindi, Khalifa M and Al-Mawali, Adhra and Akharusi, Amira and Alshukaili, Duhai and Alnasiri, Noura and Al-Awadhi, Talal and Charabi, Yassine and El Kenawy, Ahmed M},
  journal={Geospat Health},
  volume={16},
  number={1},
  year={2021}
}

@article{harris2021measuring,
  title={Measuring the exposure of {B}lack, {A}sian and other ethnic groups to {COVID}-infected neighbourhoods in {E}nglish towns and cities},
  author={Harris, Richard and Brunsdon, Chris},
  journal={Appl Spat Anal Policy},
  pages={1--26},
  year={2021},
  publisher={Springer}
}

@article{akinwumiju2022geospatial,
  title={Geospatial evaluation of {COVID-19} mortality: {I}nfluence of socio-economic status and underlying health conditions in contiguous {USA}},
  author={Akinwumiju, Akinola S and Oluwafemi, Olawale and Mohammed, Yahaya D and Mobolaji, Jacob W},
  journal={Appl Geogr},
  volume={141},
  pages={102671},
  year={2022},
  publisher={Elsevier}
}

@article{choi2021studying,
  title={Studying the social determinants of {COVID-19} in a data vacuum},
  author={Choi, Kate H and Denice, Patrick and Haan, Michael and Zajacova, Anna},
  journal={Can Rev Sociol},
  volume={58},
  number={2},
  pages={146--164},
  year={2021},
  publisher={Wiley Online Library}
}

@article{feng2022spatial,
  title={Spatial-temporal generalized additive model for modeling {COVID-19} mortality risk in {T}oronto, {C}anada},
  author={Feng, Cindy},
  journal={Spat Stat},
  volume={49},
  pages={100526},
  year={2022},
  publisher={Elsevier}
}

@article{diaz2020association,
  title={Association between the new {COVID-19} cases and air pollution with meteorological elements in nine counties of {N}ew {Y}ork state},
  author={D{\'\i}az-Avalos, Carlos and Juan, Pablo and Chaudhuri, Somnath and S{\'a}ez, Marc and Serra, Laura},
  journal={Int J Environ Res Public Health},
  volume={17},
  number={23},
  pages={9055},
  year={2020},
  publisher={MDPI}
}

@article{wang2020high,
  title={High temperature and high humidity reduce the transmission of {COVID-19}},
  author={Wang, Jingyuan and Tang, Ke and Feng, Kai and Lv, Weifeng and others},
  journal={Available at SSRN},
  volume={3551767},
  pages={2020b},
  year={2020}
}

@article{wu2020air,
  title={Air pollution and {COVID-19} mortality in the {U}nited {S}tates: {S}trengths and limitations of an ecological regression analysis},
  author={Wu, Xiao and Nethery, Rachel C and Sabath, M Benjamin and Braun, Danielle and Dominici, Francesca},
  journal={Sci Adv},
  volume={6},
  number={45},
  pages={eabd4049},
  year={2020},
  publisher={American Association for the Advancement of Science}
}

@article{mansour2021sociodemographic,
  title={Sociodemographic determinants of {COVID-19} incidence rates in {O}man: {G}eospatial modelling using multiscale geographically weighted regression ({MGWR})},
  author={Mansour, Shawky and Al Kindi, Abdullah and Al-Said, Alkhattab and Al-Said, Adham and Atkinson, Peter},
  journal={Sustain Cities Soc},
  volume={65},
  pages={102627},
  year={2021},
  publisher={Elsevier}
}

@article{kim2021covid,
  title={{COVID-19} testing, case, and death rates and spatial socio-demographics in {N}ew {Y}ork {C}ity: {A}n ecological analysis as of {J}une 2020},
  author={Kim, Byoungjun and Rundle, Andrew G and Goodwin, Alicia T Singham and Morrison, Christopher N and Branas, Charles C and El-Sadr, Wafaa and Duncan, Dustin T},
  journal={Health Place},
  volume={68},
  pages={102539},
  year={2021},
  publisher={Elsevier}
}

@article{gebhard2020impact,
  title={Impact of sex and gender on {COVID-19} outcomes in Europe},
  author={Gebhard, Catherine and Regitz-Zagrosek, Vera and Neuhauser, Hannelore K and Morgan, Rosemary and Klein, Sabra L},
  journal={Biol Sex Differ},
  volume={11},
  pages={1--13},
  year={2020},
  publisher={Springer}
}

@article{mollalo2020gis,
  title={GIS-based spatial modeling of COVID-19 incidence rate in the continental United States},
  author={Mollalo, Abolfazl and Vahedi, Behzad and Rivera, Kiara M},
  journal={Science of the total environment},
  volume={728},
  pages={138884},
  year={2020},
  publisher={Elsevier}
}

@article{castro2021spatial,
  title={Spatial dynamics of the {COVID-19} pandemic in {B}razil},
  author={Castro, RR and Santos, RSC and Sousa, GJB and Pinheiro, YT and Martins, RRIM and Pereira, MLD and Silva, RAR},
  journal={Epidemiol Infect},
  volume={149},
  year={2021},
  publisher={Cambridge University Press}
}

@article{liu2021impacts,
  title={The impacts of the built environment on the incidence rate of {COVID-19}: {A} case study of {K}ing {C}ounty, {W}ashington},
  author={Liu, Chao and Liu, Zerun and Guan, ChengHe},
  journal={Sustainable cities and society},
  volume={74},
  pages={103144},
  year={2021},
  publisher={Elsevier}
}

@article{huang2021long,
  title={Long-term exposure to air pollution and {COVID-19} incidence: a multi-country study},
  author={Huang, Guowen and Blangiardo, Marta and Brown, Patrick E and Pirani, Monica},
  journal={Spat Spatiotemporal Epidemio},
  volume={39},
  pages={100443},
  year={2021},
  publisher={Elsevier}
}

@article{sartorius2021modelling,
  title={Modelling and predicting the spatio-temporal spread of {COVID-19}, associated deaths and impact of key risk factors in {E}ngland},
  author={Sartorius, B and Lawson, AB and Pullan, RL},
  journal={Scientific reports},
  volume={11},
  number={1},
  pages={1--11},
  year={2021},
  publisher={Springer}
}

@article{cockings2011maintaining,
  title={Maintaining existing zoning systems using automated zone-design techniques: methods for creating the 2011 {C}ensus output geographies for {E}ngland and {W}ales},
  author={Cockings, Samantha and Harfoot, Andrew and Martin, David and Hornby, Duncan},
  journal={Environment and Planning A},
  volume={43},
  number={10},
  pages={2399--2418},
  year={2011},
  publisher={SAGE Publications Sage UK: London, England}
}

@article{moran1950notes,
  title={Notes on continuous stochastic phenomena},
  author={Moran, Patrick AP},
  journal={Biometrika},
  volume={37},
  number={1/2},
  pages={17--23},
  year={1950},
  publisher={JSTOR}
}

@article{ljung1978measure,
  title={On a measure of lack of fit in time series models},
  author={Ljung, Greta M and Box, George EP},
  journal={Biometrika},
  volume={65},
  number={2},
  pages={297--303},
  year={1978},
  publisher={Oxford University Press}
}

@article{haining2010inference,
  title={Inference from ecological models: estimating the relative risk of stroke from air pollution exposure using small area data},
  author={Haining, Robert and Li, Guangquan and Maheswaran, Ravi and Blangiardo, Marta and Law, Jane and Best, Nicky and Richardson, Sylvia},
  journal={Spat Spatiotemporal Epidemio},
  volume={1},
  number={2-3},
  pages={123--131},
  year={2010},
  publisher={Elsevier}
}

@book{rothman2008modern,
  title={Modern epidemiology},
  author={Rothman, Kenneth J and Greenland, Sander and Lash, Timothy L and others},
  volume={3},
  year={2008},
  publisher={Wolters Kluwer Health/Lippincott Williams \& Wilkins Philadelphia}
}

@article{lindgren2011explicit,
  title={An explicit link between {G}aussian fields and {G}aussian {M}arkov random fields: the stochastic partial differential equation approach},
  author={Lindgren, Finn and Rue, H{\aa}vard and Lindstr{\"o}m, Johan},
  journal={J R Stat Soc Series B Stat Methodo},
  volume={73},
  number={4},
  pages={423--498},
  year={2011},
  publisher={Wiley Online Library}
}

@article{rue2009approximate,
  title={Approximate {B}ayesian inference for latent {G}aussian models by using integrated nested {L}aplace approximations},
  author={Rue, H{\aa}vard and Martino, Sara and Chopin, Nicolas},
  journal={J R Stat Soc Series B Stat Methodo},
  volume={71},
  number={2},
  pages={319--392},
  year={2009},
  publisher={Wiley Online Library}
}

@book{team2013r,
  title={R: {A} language and environment for statistical computing},
  author={{R Core Team}},
  year={2013},
  publisher={Vienna, Austria}
}

@book{blangiardo2015spatial,
  title={Spatial and spatio-temporal {B}ayesian models with {R-INLA}},
  author={Blangiardo, Marta and Cameletti, Michela},
  year={2015},
  publisher={John Wiley \& Sons}
}

@book{rue2005gaussian,
  title={Gaussian {M}arkov random fields: theory and applications},
  author={Rue, Havard and Held, Leonhard},
  year={2005},
  publisher={Chapman and Hall/CRC}
}

@article{fuglstad2019constructing,
  title={Constructing priors that penalize the complexity of {G}aussian random fields},
  author={Fuglstad, Geir-Arne and Simpson, Daniel and Lindgren, Finn and Rue, H{\aa}vard},
  journal={J Am Stat Assoc},
  volume={114},
  number={525},
  pages={445--452},
  year={2019},
  publisher={Taylor \& Francis}
}

@article{leffler2020association,
  title={Association of country-wide coronavirus mortality with demographics, testing, lockdowns, and public wearing of masks},
  author={Leffler, Christopher T and Ing, Edsel and Lykins, Joseph D and Hogan, Matthew C and McKeown, Craig A and Grzybowski, Andrzej},
  journal={Am J Trop Med Hyg},
  volume={103},
  number={6},
  pages={2400},
  year={2020},
  publisher={The American Society of Tropical Medicine and Hygiene}
}

@article{takagi2021meta,
  title={Meta-regression of {COVID-19} prevalence/fatality on socioeconomic characteristics of data from top 50 {US} large cities},
  author={Takagi, Hisato and Kuno, Toshiki and Yokoyama, Yujiro and Ueyama, Hiroki and Matsushiro, Takuya and Hari, Yosuke and Ando, Tomo},
  journal={J Med Virol},
  volume={93},
  number={2},
  pages={595},
  year={2021},
  publisher={Wiley-Blackwell}
}

@article{davies2020estimated,
  title={Estimated transmissibility and severity of novel {SARS}-{C}o{V}-2 {V}ariant of {C}oncern 202012/01 in {E}ngland},
  author={Davies, Nicholas G and Abbott, Sam and Barnard, Rosanna C and Jarvis, Christopher I and Kucharski, Adam J and Munday, James and Pearson, Carl AB and Russell, Timothy W and Tully, Damien C and Washburne, Alex D and others},
  journal={medRxiv},
  pages={2020--12},
  year={2020},
  publisher={Cold Spring Harbor Laboratory Press}
}

@incollection{batty2021london,
  title={London in lockdown: {M}obility in the pandemic city},
  author={Batty, Michael and Murcio, Roberto and Iacopini, Iacopo and Vanhoof, Maarten and Milton, Richard},
  booktitle={COVID-19 pandemic, Geospatial Information, and Community Resilience},
  pages={229--244},
  year={2021},
  publisher={CRC Press}
}

@article{mecenas2020effects,
  title={Effects of temperature and humidity on the spread of {COVID-19}: {A} systematic review},
  author={Mecenas, Paulo and Bastos, Renata Travassos da Rosa Moreira and Vallinoto, Antonio Carlos Ros{\'a}rio and Normando, David},
  journal={PLoS One},
  volume={15},
  number={9},
  pages={e0238339},
  year={2020},
  publisher={Public Library of Science San Francisco, CA USA}
}

@article{wang2021impact,
  title={Impact of temperature and relative humidity on the transmission of {COVID-19}: a modelling study in {C}hina and the {U}nited {S}tates},
  author={Wang, Jingyuan and Tang, Ke and Feng, Kai and Lin, Xin and Lv, Weifeng and Chen, Kun and Wang, Fei},
  journal={BMJ Open},
  volume={11},
  number={2},
  pages={e043863},
  year={2021},
  publisher={British Medical Journal Publishing Group}
}

@article{maiti2021exploring,
  title={Exploring spatiotemporal effects of the driving factors on {COVID-19} incidences in the contiguous {U}nited {S}tates},
  author={Maiti, Arabinda and Zhang, Qi and Sannigrahi, Srikanta and Pramanik, Suvamoy and Chakraborti, Suman and Cerda, Artemi and Pilla, Francesco},
  journal={Sustain Cities Soc},
  volume={68},
  pages={102784},
  year={2021},
  publisher={Elsevier}
}

@article{zhai2021american,
  title={American inequality meets {COVID-19}: {U}neven spread of the disease across communities},
  author={Zhai, Wei and Liu, Mengyang and Fu, Xinyu and Peng, Zhong-Ren},
  journal={Ann Am Assoc Geogr},
  volume={111},
  number={7},
  pages={2023--2043},
  year={2021},
  publisher={Taylor \& Francis}
}

@article{fattorini2020role,
  title={Role of the chronic air pollution levels in the {C}ovid-19 outbreak risk in {I}taly},
  author={Fattorini, Daniele and Regoli, Francesco},
  journal={Environ Pollut},
  volume={264},
  pages={114732},
  year={2020},
  publisher={Elsevier}
}

@article{comunian2020air,
  title={Air pollution and {COVID-19}: the role of particulate matter in the spread and increase of {COVID-19}’s morbidity and mortality},
  author={Comunian, Silvia and Dongo, Dario and Milani, Chiara and Palestini, Paola},
  journal={Int J Environ Res Public Health},
  volume={17},
  number={12},
  pages={4487},
  year={2020},
  publisher={MDPI}
}

@article{semczuk2021association,
  title={Association between air pollution and {COVID-19} mortality and morbidity},
  author={Semczuk-Kaczmarek, Karolina and Rys-Czaporowska, Anna and Sierdzinski, Janusz and Kaczmarek, Lukasz Dominik and Szymanski, Filip Marcin and Platek, Anna Edyta},
  journal={Intern Emerg Med},
  pages={1--7},
  year={2021},
  publisher={Springer}
}

@article{wakefield2001statistical,
  title={A statistical framework for ecological and aggregate studies},
  author={Wakefield, Jonathan and Salway, Ruth},
  journal={J R Stat Soc Ser A Stat Soc},
  volume={164},
  number={1},
  pages={119--137},
  year={2001},
  publisher={Wiley Online Library}
}

@article{wintle2003use,
  title={The use of {B}ayesian model averaging to better represent uncertainty in ecological models},
  author={Wintle, Brendan A and McCarthy, Michel A and Volinsky, Chris T and Kavanagh, Rodney P},
  journal={Conserv Biol},
  volume={17},
  number={6},
  pages={1579--1590},
  year={2003},
  publisher={Wiley Online Library}
}

@article{cressie2009accounting,
  title={Accounting for uncertainty in ecological analysis: the strengths and limitations of hierarchical statistical modeling},
  author={Cressie, Noel and Calder, Catherine A and Clark, James S and Hoef, Jay M Ver and Wikle, Christopher K},
  journal={Ecol Appl},
  volume={19},
  number={3},
  pages={553--570},
  year={2009},
  publisher={Wiley Online Library}
}

@article{wu2021exploration,
  title={Exploration of spatial-temporal varying impacts on {COVID-19} cumulative case in {T}exas using geographically weighted regression ({GWR})},
  author={Wu, Xiu and Zhang, Jinting},
  journal={Environmental Science and Pollution Research},
  volume={28},
  pages={43732--43746},
  year={2021},
  publisher={Springer}
}

@article{zhang2021space,
  title={Space-time cluster’s detection and geographical weighted regression analysis of {COVID-19} mortality on {T}exas counties},
  author={Zhang, Jinting and Wu, Xiu and Chow, T Edwin},
  journal={International Journal of Environmental Research and Public Health},
  volume={18},
  number={11},
  pages={5541},
  year={2021},
  publisher={MDPI}
}

@article{kulu2021infection,
  title={Infection rates from {C}ovid-19 in {G}reat {B}ritain by geographical units: A model-based estimation from mortality data},
  author={Kulu, Hill and Dorey, Peter},
  journal={Health \& place},
  volume={67},
  pages={102460},
  year={2021},
  publisher={Elsevier}
}

@article{jackson2021spatial,
  title={Spatial disparities of {COVID-19} cases and fatalities in {U}nited {S}tates counties},
  author={Jackson, Sarah L and Derakhshan, Sahar and Blackwood, Leah and Lee, Logan and Huang, Qian and Habets, Margot and Cutter, Susan L},
  journal={International Journal of Environmental Research and Public Health},
  volume={18},
  number={16},
  pages={8259},
  year={2021},
  publisher={MDPI}
}

@article{liu2021spatiotemporal,
  title={Spatiotemporal analysis of {COVID-19} outbreaks in {W}uhan, {C}hina},
  author={Liu, Wei and Wang, Dongming and Hua, Shuiqiong and Xie, Cong and Wang, Bin and Qiu, Weihong and Xu, Tao and Ye, Zi and Yu, Linling and Yang, Meng and others},
  journal={Sci. Rep},
  volume={11},
  number={1},
  pages={1--9},
  year={2021},
  publisher={Springer}
}

@article{sarkar2021spatial,
  title={Spatial modeling of {COVID-19} transmission in Bangladesh},
  author={Sarkar, Showmitra Kumar and Ekram, Khondaker Mohammed Mohiuddin and Das, Palash Chandra},
  journal={Spat. Inf. Res},
  pages={1--12},
  year={2021},
  publisher={Springer}
}

@article{martines2021detecting,
  title={Detecting space--time clusters of {COVID-19} in {B}razil: mortality, inequality, socioeconomic vulnerability, and the relative risk of the disease in {B}razilian municipalities},
  author={Martines, Marcos R and Ferreira, Ricardo Vicente and Toppa, Rog{\'e}rio Hartung and Assun{\c{c}}{\~a}o, LM and Desjardins, Michael R and Delmelle, Eric M},
  journal={J Geogr Syst}, 
  volume={23},
  pages={7--36},
  year={2021},
  publisher={Springer}
}

@article{mendy2021air,
  title={Air pollution and the pandemic: {L}ong-term {PM}2. 5 exposure and disease severity in {COVID-19} patients},
  author={Mendy, Angelico and Wu, Xiao and Keller, Jason L and Fassler, Cecily S and Apewokin, Senu and Mersha, Tesfaye B and Xie, Changchun and Pinney, Susan M},
  journal={Respirology},
  volume={26},
  number={12},
  pages={1181--1187},
  year={2021},
  publisher={Wiley Online Library}
}

@article{mathur2021ethnic,
  title={Ethnic differences in {SARS}-{C}o{V}-2 infection and {COVID-19}-related hospitalisation, intensive care unit admission, and death in 17 million adults in {E}ngland: an observational cohort study using the {O}pen{SAFELY} platform},
  author={Mathur, Rohini and Rentsch, Christopher T and Morton, Caroline E and Hulme, William J and Schultze, Anna and MacKenna, Brian and Eggo, Rosalind M and Bhaskaran, Krishnan and Wong, Angel YS and Williamson, Elizabeth J and others},
  journal={Lancet},
  volume={397},
  number={10286},
  pages={1711--1724},
  year={2021},
  publisher={Elsevier}
}

@book{williams2006gaussian,
  title={Gaussian processes for machine learning},
  author={Williams, Christopher KI and Rasmussen, Carl Edward},
  volume={2},
  number={3},
  year={2006},
  publisher={MIT press Cambridge, MA}
}

@article{green2023occupational,
  title={Occupational inequalities in the prevalence of {COVID-19}: {A} longitudinal observational study of {E}ngland, {A}ugust 2020 to {J}anuary 2021},
  author={Green, Mark A and Semple, Malcolm G},
  journal={PLoS One},
  volume={18},
  number={4},
  pages={e0283119},
  year={2023},
  publisher={Public Library of Science San Francisco, CA USA}
}

@article{richardson2022association,
  title={Association of race/ethnicity with mortality in patients hospitalized with {COVID-19}},
  author={Richardson, Safiya and Martinez, Johanna and Hirsch, Jamie S and Cerise, Jane and Lesser, Martin and Roswell, Robert O and Davidson, Karina W and Northwell Health COVID-19 Research Consortium},
  journal={PLoS One},
  volume={17},
  number={8},
  pages={e0267505},
  year={2022},
  publisher={Public Library of Science San Francisco, CA USA}
}

@Book{hadley2016ggplot2,
  author = {Hadley Wickham},
  title = {ggplot2: {E}legant {G}raphics for {D}ata {A}nalysis},
  publisher = {Springer-Verlag New York},
  year = {2016},
  isbn = {978-3-319-24277-4},
  url = {https://ggplot2.tidyverse.org},
}

@article{neuwirth2014colorbrewer,
  title={Color{B}rewer palettes},
  author={Neuwirth, Erich and Brewer, R Color},
  journal={R package version},
  volume={1},
  pages={4},
  year={2014}
}

@article{bivand2019rgdal,
  title={rgdal: {B}indings for the geospatial data abstraction library},
  author={Bivand, Roger and Keitt, Tim and Rowlingson, Barry and Pebesma, EDZER and Sumner, Michael and Hijmans, Robert and others},
  journal={R package version},
  volume={1},
  number={3},
  year={2019}
}

@article{wickham2015dplyr,
  title={dplyr: {A} grammar of data manipulation},
  author={Wickham, Hadley and Fran{\c{c}}ois, Romain and Henry, Lionel and M{\"u}ller, Kirill and others},
  journal={R package version 0.4},
  volume={3},
  pages={p156},
  year={2015}
}

@article{cheng2019leaflet,
  title={Leaflet: create interactive web maps with the {J}ava{S}cript'{L}eaflet'},
  author={Cheng, Joe},
  journal={library},
  year={2019}
}

@Article{spde2022,
    author = {{Roger Bivand}},
    title = {R Packages for {A}nalyzing {S}patial {D}ata: {A} {C}omparative {C}ase
      {S}tudy with {A}real {D}ata},
    journal = {Geogr Anal},
    year = {2022},
    volume = {54},
    number = {3},
    pages = {488-518},
    doi = {10.1111/gean.12319},
  }

@article{tamrakar2021district,
  title={District level correlates of {COVID}-19 pandemic in {I}ndia during {M}arch-{O}ctober 2020},
  author={Tamrakar, Vandana and Srivastava, Ankita and Saikia, Nandita and Parmar, Mukesh C and Shukla, Sudheer Kumar and Shabnam, Shewli and Boro, Bandita and Saha, Apala and Debbarma, Benjamin},
  journal={PLoS One},
  volume={16},
  number={9},
  pages={e0257533},
  year={2021},
  publisher={Public Library of Science San Francisco, CA USA}
}

@article{wong2020spreading,
  title={Spreading of {COVID}-19: {D}ensity matters},
  author={Wong, David WS and Li, Yun},
  journal={PLoS One},
  volume={15},
  number={12},
  pages={e0242398},
  year={2020},
  publisher={Public Library of Science San Francisco, CA USA}
}

@article{muegge2023national,
  title={National lockdowns in {E}ngland: {T}he same restrictions for all, but do the impacts on {COVID-19} mortality risks vary geographically?},
  author={Muegge, Robin and Dean, Nema and Jack, Eilidh and Lee, Duncan},
  journal={Spat Spatiotemporal Epidemio},
  volume={44},
  pages={100559},
  year={2023},
  publisher={Elsevier}
}

@book{menard2002applied,
  title={Applied logistic regression analysis},
  author={Menard, Scott},
  number={106},
  year={2002},
  publisher={Sage}
}

@article{goutte2020role,
  title={The role of economic structural factors in determining pandemic mortality rates: {E}vidence from the {COVID-19} outbreak in France},
  author={Goutte, St{\'e}phane and P{\'e}ran, Thomas and Porcher, Thomas},
  journal={Res. Int. Bus. Finance},
  volume={54},
  pages={101281},
  year={2020},
  publisher={Elsevier}
}

@article{nazia2022identifying,
  title={Identifying spatiotemporal patterns of COVID-19 transmissions and the drivers of the patterns in Toronto: a Bayesian hierarchical spatiotemporal modelling},
  author={Nazia, Nushrat and Law, Jane and Butt, Zahid Ahmad},
  journal={Sci. Rep}, 
  volume={12},
  number={1},
  pages={9369},
  year={2022},
  publisher={Nature Publishing Group UK London}
}

@article{konstantinoudis2021long,
  title={Long-term exposure to air-pollution and {COVID-19} mortality in {E}ngland: a hierarchical spatial analysis},
  author={Konstantinoudis, Garyfallos and Padellini, Tullia and Bennett, James and Davies, Bethan and Ezzati, Majid and Blangiardo, Marta},
  journal={Environment international},
  volume={146},
  pages={106316},
  year={2021},
  publisher={Elsevier}
}

@book{gelman1995bayesian,
  title={Bayesian data analysis},
  author={Gelman, Andrew and Carlin, John B and Stern, Hal S and Rubin, Donald B},
  year={1995},
  publisher={Chapman and Hall/CRC}
}

@article{hartigan1979algorithm,
  title={Algorithm {AS 136}: {A} k-means clustering algorithm},
  author={Hartigan, John A and Wong, Manchek A},
  journal={Journal of the Royal Statistical Society. Series C (Applied Statistics)},
  volume={28},
  number={1},
  pages={100--108},
  year={1979},
  publisher={JSTOR}
}

@article{lee2021quantifying,
  title={Quantifying the small-area spatio-temporal dynamics of the {C}ovid-19 pandemic in {S}cotland during a period with limited testing capacity},
  author={Lee, Duncan and Robertson, Chris and Marques, Diogo},
  journal={Spat Stat},
  pages={100508},
  year={2021},
  publisher={Elsevier}
}

\end{document}